\documentclass[traditabstract,letter]{aa}  
\usepackage{graphicx,amsmath}

\begin{document}
\title{
GMRT radio observations of the transiting extrasolar planet HD189733\,b 
at 244 and 614~MHz\thanks{Data for this observations can be retrieved electronically 
on the GMRT archive server {\tt http://ncra.tifr.res.in/\symbol{126}gmrtarchive} 
and by request to {\tt archive@gmrt.ncra.tifr.res.in}.}
    }

   \author{
 A.~Lecavelier des Etangs\inst{1,2}
 \and
 S.~K.~Sirothia\inst{3}
 \and
 Gopal-Krishna\inst{3}       
 \and
 P.~Zarka\inst{4}
   }
   

\titlerunning{GMRT radio observations of the transiting extrasolar planet HD189733\,b}


   \institute{
   CNRS, UMR 7095, 
   Institut d'Astrophysique de Paris, 
   98$^{\rm bis}$ boulevard Arago, F-75014 Paris, France
   \and
   UPMC Univ. Paris 6, UMR 7095, 
   Institut d'Astrophysique de Paris, 
   98$^{\rm bis}$ boulevard Arago, F-75014 Paris, France
   \and
   National Centre for Radio Astrophysics, TIFR, 
   Post Bag 3, Pune University Campus, Pune 411007, India
   \and
   LESIA, Observatoire de Paris, CNRS, UPMC, Universit\'e Paris Diderot, 
   5 Place Jules Janssen, 92190 Meudon, France
}   
   \date{} 
 
  \abstract
{
%

We report a sensitive search for meter-wavelength emission at 244 and 614~MHz 
from HD189733\,b, the nearest known extrasolar transiting planet of `hot-Jupiter' 
type. To discriminate any planetary emission from possible
stellar or background contributions, 
we observed the system for 7.7~hours 
encompassing the planet's eclipse behind the host star. These GMRT observations 
provide very low (3$\sigma$) upper limits of
2~mJy at 244~MHz and 160~$\mu$Jy at 614~MHz. These limits are, respectively, 
about 40 and 500~times deeper than those reported recently at 
a nearby frequency of 340 MHz. Possible explanations of our non-detection 
include: (1) the Earth being outside the planet's emission beam; 
(2) its highly variable emission with more rapid flaring than 
the temporal sampling in our observations; (3) the planetary 
emission being intrinsically too weak; or more likely, (4) 
the emission being predominantly at lower frequencies because of a weak
planetary magnetic field. We briefly discuss these possibilities and the
constraints on this exo-planetary system environment.  }


   \maketitle
%

\section{Introduction}
\label{Introduction}

The detection of radio emission from an extrasolar planet would be a 
major step in the characterization of these planets and their environment, 
possibly providing information about the planetary magnetic field and the 
interaction of the planet with the stellar magnetic field and corona.
In parallel with theoretical estimates for radio emission from a large number 
of extrasolar planets ({\it e.g.,} Grie\ss meier et al.\ 2007), 
searches for decameter- and meter-wavelength radio emission from a few 
extrasolar planets have been undertaken ({\it e.g.,} 
Bastian et al.\ 2000; Ryabov et al.\ 2004; Winterhalter et al.\ 2005; 
George \& Stevens 2007; Lazio \& Farrell 2007).
Radio detections seem currently feasible only provided that the planets
are 10$^3$ to 10$^4$ times stronger emitters than Jupiter. However, the 
extreme conditions of ``hot-Jupiters'' could make this happen
because of the extreme incident Poynting flux (Zarka 2007),
justifying the radio-magnetic scaling law proposed in Zarka et al.\ (2001).
At any rate, all the 
estimates of cyclotron maser decametric emission involve several unknowns, 
{\it e.g.}, stellar winds, coronal density, and stellar and planetary 
magnetic fields.

So far only non-detections have been reported, the telescopes used being 
UTR (10-30\,MHz, $\sigma$$\sim$1.6\,Jy), VLA (74\,MHz, $\sigma$$\sim$50\,mJy), 
and GMRT (150\,MHz, $\sigma$$\sim$10\,mJy; see review in Lazio et al.\ 2009). 
The principal contributors to the 
noise level at these low frequencies are the sky background, 
radio frequency interference, and ionospheric scintillations, which 
distort the incoming signal and increase the noise.
Hence, 
interferometric observations of high sensitivity and resolution hold 
considerable promise for improvement. In particular, the Giant Metrewave Radio
Telescope (GMRT), a 30-km baseline array consisting of 30 dishes of 45
metre diameter each (Swarup 1990), appears to be the telescope of choice.
We report the first GMRT search targeted at the planet HD189733\,b,
which is one of the best candidates among the known ``hot-Jupiter'' extrasolar planets
(Sect.~\ref{HD189733b}). Our search is over an order-of-magnitude deeper 
than the recently reported search for meter-wavelength emission from this
system based on single-dish observations at 307-347\,MHz (Smith et al.\ 2009).

\section{The target planet: HD\,189733\,b}
\label{HD189733b}

Located just 19.3~parsecs away, HD189733\,b is one of the most prominent 
extrasolar planets known (Bouchy et al.\ 2005). With a semi-major axis 
of 0.03~AU and an orbital period of 2.2 days, it belongs to the class of 
``very hot-Jupiters". More importantly, since this planet is seen to
transit its parent star, the planetary transits and eclipses can be
used to probe the planet's atmosphere and environment 
({\it e.g.}, Charbonneau et al.\ 2008, D\'esert et al.\ 2009).    

HD189733b orbits a small and bright main-sequence K-type star, and shows a 
transit occultation depth of $\approx$2.5\% at optical wavelengths
(Pont et al.\ 2007). The
planet has a mass $M_p$=1.13~Jupiter masses ($M_{\rm Jup}$) and a
radius $R_p$=1.16~Jupiter radii ($R_{\rm Jup}$) in the visible (Bakos
et al.\ 2006; Winn et al.\ 2007). The short period of the planet (2.21858 
days) has been measured precisely (H\'ebrard \& Lecavelier des 
Etangs 2006; Knutson et al.\ 2009). 
Spectropolarimetry has measured the strength and topology 
of the stellar magnetic field, which reaches up to 40~G (Moutou et al.\ 2007).
Sodium has been detected in the planet's atmosphere
by ground-based observations (Redfield et al.\ 2008). 
Using the `Advanced Camera for Survey' aboard the Hubble Space Telescope 
(HST), Pont et al.\ (2008) detected atmospheric haze, which is interpreted as Mie
scattering by small particles (Lecavelier des Etangs et al.\ 2008). 
CO molecules have been tentatively proposed to explain the excess 
absorption seen at 4.5~$\mu$m (Charbonneau et al.\ 2008; D\'esert et al.\ 2009). 
Absorption in Lyman-$\alpha$ observed with the HST/ACS is explained in terms
of atomic hydrogen escaping from the planet's exosphere at a rate 
of 10$^7$-10$^{11}$~g/s (Lecavelier des Etangs et al., in preparation). 

The atmosphere and environment of transiting planets can also be
studied using the planetary eclipse techniques. The principle is to 
subtract the signal received when the planet 
is hidden behind the star, from observations made before and after this
eclipse. This allows reliable 
extraction of the planetary emission. This technique has
enabled detection of thermal infrared emission from the extrasolar planets 
HD\,209458\,b and 
TReS-1 (Charbonneau et al.\ 2005; Deming et al.\ 2005).  Using Spitzer 
spectroscopy of planetary eclipses, the infrared spectra of HD189733\,b 
atmosphere have revealed signatures of H$_2$O absorption and possibly 
weather-like variations in the atmospheric conditions (Grillmair et al.\ 2009).

We have developed a similar strategy by monitoring the radio flux from 
the HD\,189733\,b system before, 
during, and after a planetary eclipse. By comparing 
the radio flux levels, we can thus discriminate between any radio emission 
contributed by the planet and the star (or any other background source within 
the synthesized beam). 

\section{Observations and data analysis}
\label{HD189733_esp:sec:obs}

Using GMRT, we performed 
simultaneous dual-frequency observations of the HD189733b field at
244~MHz and 614~MHz on 2008 August 14. The phase centre was set at the star's position of $\alpha_0$=20h00m43.7s, $\delta_0$=+22\degr42\arcmin39\arcsec (J2000).
The observations started at 13h50m and finished at 22h20m (UT), covering 
the full passage of the target in the visibility window of the sky and the 
planet's eclipse behind the star, which took place between 15h55m and
17h43m~(UT). 
At 244 MHz, the receiver bandwidth was 5.6~MHz (LL polarization only) 
and at 614 MHz the bandwidth was 32~MHz (RR polarization only).
At each frequency, the visibility integration time was 16.78~seconds 
and the total observation time was 7.71~hours, on both the primary field 
and the calibration sources.
Because of a temporary system failure, roughly in the 
middle of the run, most of the antennas could not be used for $\sim$1.5 hours
from 16h36m to 18h10m (UT). 
Although this observing `gap' substantially overlapped with the 
planet's occultation behind the star, this problem should not
affect our main conclusions.

At both frequencies, 3C48 was observed as the primary flux density 
and bandpass calibrator, for a total of 0.27~hours.
The source 2052+365 was chosen as a phase calibrator and observed repeatedly 
for a total of 0.8~hours.
The total acquisition time on the target field was 6.45~hours.
 
\label{data_analysis}

\begin{figure*}[htb]
\begin{center}
\hbox{
 \includegraphics[angle=0,width=\columnwidth, viewport=0 210 594 625,clip]
 {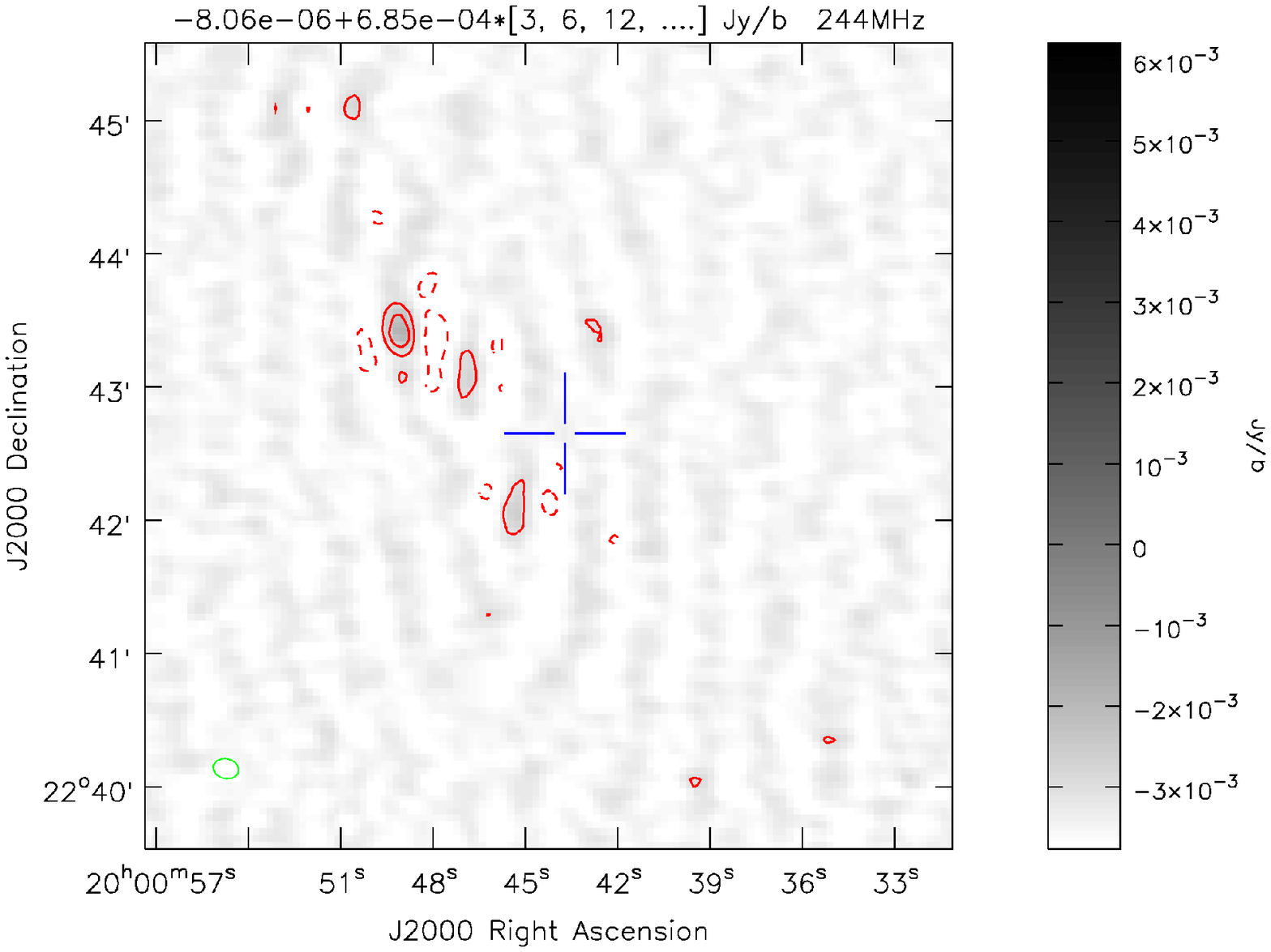}
 \includegraphics[angle=0,width=\columnwidth, viewport=0 210 594 625,clip]
 {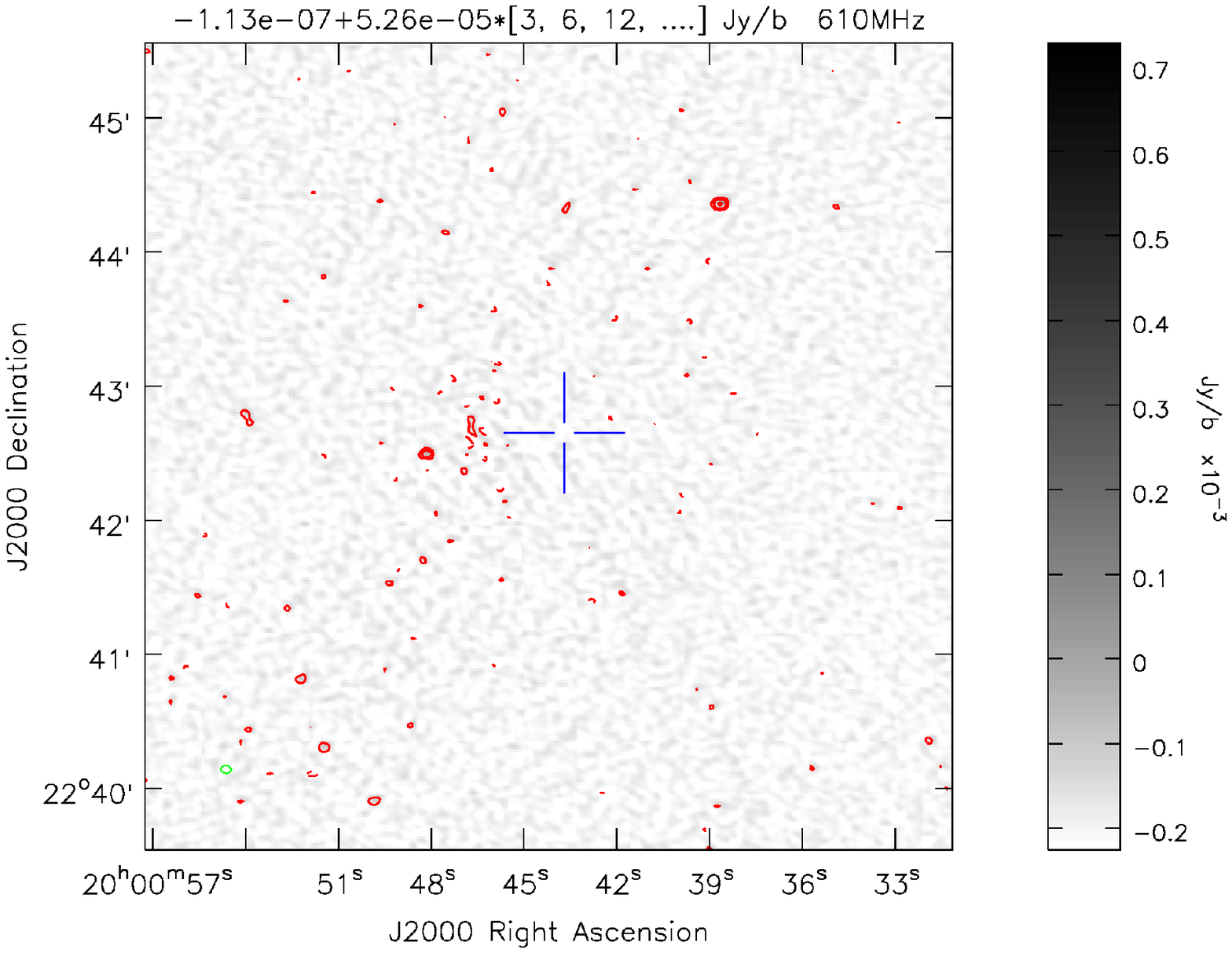}
}
\caption{GMRT image of HD189733 field at 244 MHz (left panel) and 614~MHz (right panel). The ellipse in the lower left corner of each image shows the half power beamwidth (11.54\arcsec$\times$8.82\arcsec, 77$^\circ$.6 at 244 MHz and 4.76\arcsec$\times$3.45\arcsec, 82$^\circ$.5 at 614 MHz).  The contour levels given at the top of the images are in units of Jy beam$^{-1}$ and
are defined as mean$+$rms$\times$(n) where n is an integer. Negative contours appear as dashed lines.}
\label{HD189733_esp:fig:244_614}
\end{center}
\end{figure*}

\begin{figure*}[htb]
\begin{center}
\hbox{
 \includegraphics[angle=-90,width=\columnwidth,
 viewport=50 00 505 770, clip]{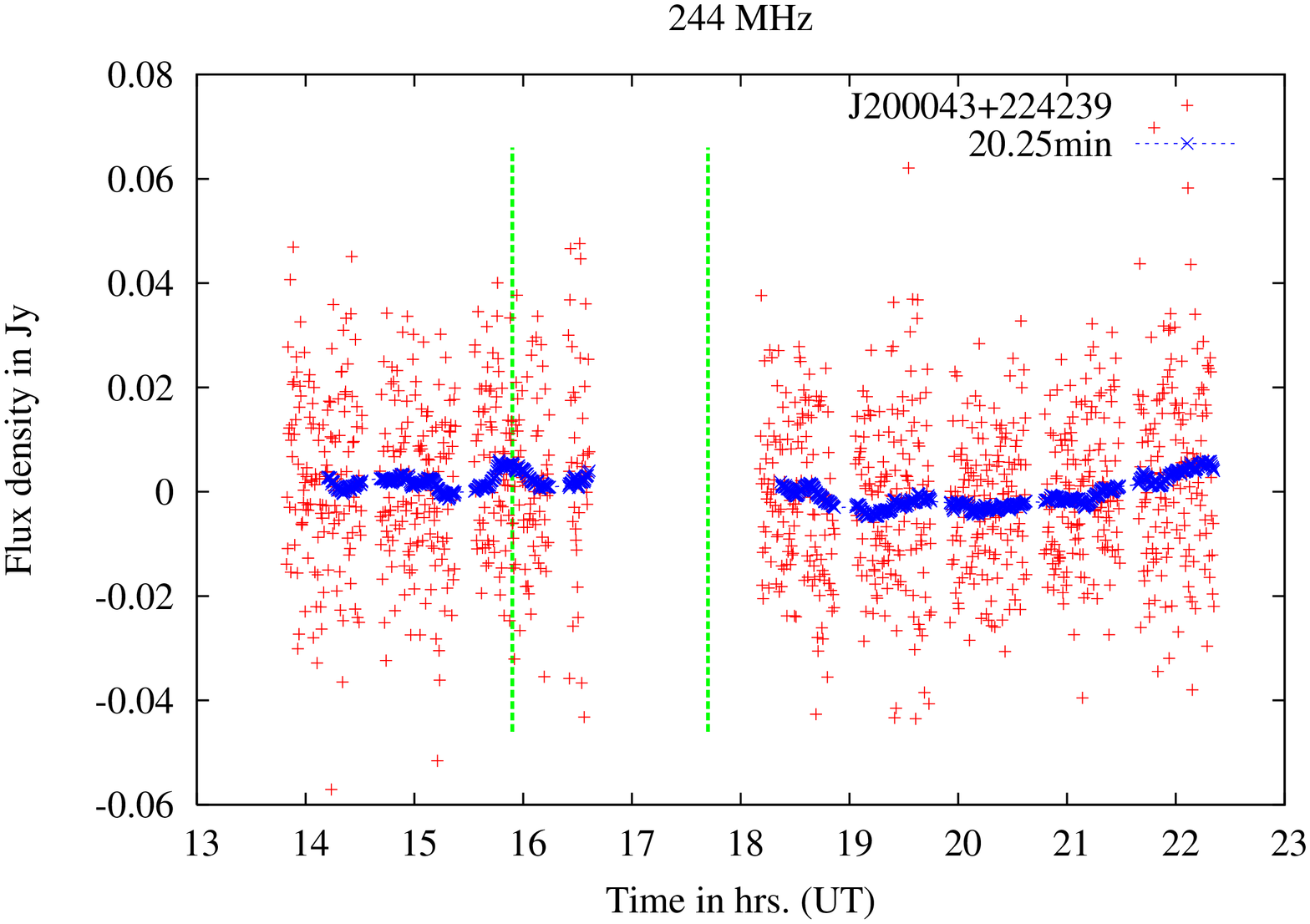}
 \includegraphics[angle=-90,width=\columnwidth,
 viewport=50 00 505 770, clip]{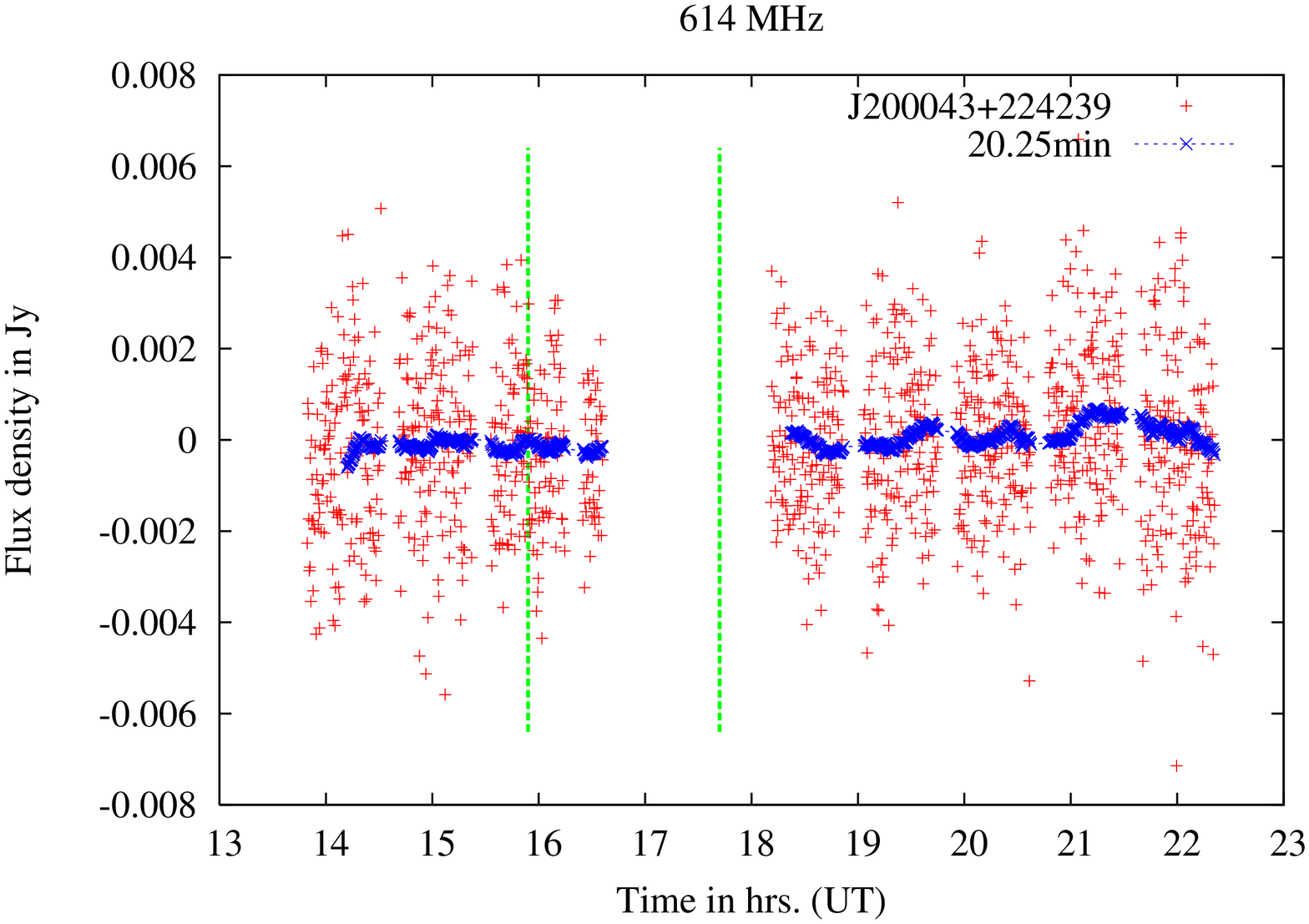}
}
\caption{Time series of the flux density measured at 244 MHz (left panel) and 
614~MHz (right panel) in the direction of HD189733 (red crosses), for a time sampling of 16.78 sec. The 
vertical dotted green lines indicate the beginning and the end of the planet's
eclipse behind the star. The thick (blue) curves show the flux densities averaged in
20.25 minute bins.}
\label{mv_ts_244_614}
\end{center}
\end{figure*}

The data reduction was completed mainly using the {\tt AIPS++} package 
(version: 1.9, build \#1556).
After applying bandpass 
corrections using 3C48, gain and phase variations were quantified and
used for the flux density, bandpass, gain, and phase calibration of the
target field data;  for 3C48, we assumed flux densities 
of 51.16~Jy and 29.305~Jy at 244~MHz and 614~MHz, respectively.

While calibrating the data, bad data points were flagged at various stages. 
The data for antennas with relatively large errors in antenna-based gain 
solutions were examined and flagged over certain time ranges. Some 
baselines were flagged, based on closure errors on the bandpass calibrator. 
Channel and time-based flagging of the data points corrupted by radio frequency 
interference (RFI) was done by applying a median filter with a $6\sigma$ threshold.
Residual errors above $5\sigma$ were also flagged after a few rounds of 
imaging and self-calibration. The system temperature ($T_{sys}$) was 
found to vary among the antennas, and also with 
the ambient temperature and elevation (Sirothia 
2009). In the absence of regular $T_{sys}$ measurements for GMRT antennas, 
this correction was estimated from the residuals of calibrated data with 
respect to the model data. The corrections were then applied to the data. 
The final image was made after several rounds of phase self-calibration, 
and one round of amplitude self-calibration, where the data were normalized 
by the median gain found for the entire data. 
The final image was also corrected for the primary beam shape taken to be a 
Gaussian 
with FWHM of 117.0\arcmin\ and 42.7\arcmin\ at the reference frequencies 
of 235~MHz and 610~MHz, respectively.
The final images have median RMS noises of about 470 and 39~$\mu$Jy per beam, 
at 244~MHz and 614~MHz, respectively. 
For the central $5^\prime\times5^\prime$ region shown in Fig.~\ref{HD189733_esp:fig:244_614}, 
we have $\sigma$=685\,$\mu$Jy at 244~MHz and $\sigma$=53\,$\mu$Jy at 614~MHz.
 
After the final imaging stage, light curves were generated
for a synthesized-beam size region centered on the coordinates 
of the star and two randomly chosen control directions a few arc minutes offset from the star. 
To obtain these light curves, we calculated model visibilities using the sources detected in the entire field-of-view excluding a synthesized beam-wide region centered 
on the desired location ($\alpha_0, \delta_0$); we then subtracted this model out from the final calibrated visibility data. The residual visibility data (RVD) were 
then phase-centered on $\alpha_0, \delta_0$ and averaged for desired time bins to
generate the light curves. Figure~\ref{mv_ts_244_614} shows the resulting light curve for the star's position.

\section{Results and discussion}
\label{HD189733_esp:sec:discussion}

\subsection{Results}
\label{HD189733_esp:sec:result}

For the stellar+planetary emission, the GMRT images 
(Fig.~\ref{HD189733_esp:fig:244_614})
provide 3$\sigma$ upper limits of 2~mJy 
at 244~MHz and 160~$\mu$Jy at 610~MHz.
We note that the 244~MHz
map shows a chain of peaks roughly along position angle
$PA \sim 35^\circ$, but offset eastward from the star's position by about four
beam widths. These peaks appear to be artefacts from the residual phase
errors dominated by the bright extended planetary nebula M27 in the
field-of-view. The upper limit of 2\,mJy 
(3$\sigma$ for the $5^\prime\times5^\prime$ field) is 2$\sigma$ for the 
$1^\prime\times5^\prime$ rectangular strip oriented at $PA \sim 35^\circ$ 
that encompasses the chain of artifacts ($\sigma$=1.05\,mJy).

Using the light curve, we also searched for a planet's emission eclipse or flares. 
With average flux densities before, during, and after the secondary transit of 
1.8$\pm$0.8~mJy, 2.7$\pm$1.7~mJy, and -0.4$\pm$0.6~mJy at 244~MHz, and -0.16$\pm$0.09~mJy, -0.38$\pm$0.15~mJy, and 0.07$\pm$0.07~mJy at 614~MHz, respectively, 
we did not detect an eclipse signature.
Using the running averages of the light curve, obtained for 
intervals between 1 and 30~minutes, 
no emission flare events were found at either frequency (Fig.~\ref{mv_ts_244_614}).
We therefore conclude a non-detection of planetary radio emission. 
We note that the planet's eclipse times plotted in Fig.~\ref{mv_ts_244_614} 
(green lines) 
are calculated using the size of the planet in the optical. If radio emission 
originates in a region larger than the planet, the ingress could happen earlier 
and the egress later. However, emission at frequencies above 200~MHz is expected to 
originate in the very inner regions of the planet magnetosphere, 
and the eclipse times are expected to be similar to that of the planet itself. 

Theoretical and observational aspects of the radio emission from 
extrasolar planets are discussed {\it e.g.}, in Zarka (2007), where
the generalized concept of flow-obstacle interaction is developed.
Accordingly, the present non-detection of HD\,189733\,b may occur because : 
(1) the Earth was outside the planet's emission beam, 
at least at the time of observation, or (2) the emission is highly variable 
with flares much faster than the temporal sampling of our 
observations, or (3) the planetary emission was simply too weak 
intrinsically, or, perhaps more likely (4) the planetary emission peaks at 
frequencies much lower than 200~MHz, because of a weak planetary magnetic field. 

\subsection{Emission beam}

Cyclotron maser emission is rather confined within a beam solid angle.
It is therefore possible that even strong emission remains unobservable 
from the Earth. 
For instance, if we assume that the magnetic axis is not too misaligned
with the rotation axis and that emission is produced at the highest cyclotron 
frequencies (i.e., very near the planetary poles), then the typical beaming
angle of $\sim$50\degr-60\degr\ may account for the non-detection. 
However, in case of a magnetic axis tilted with respect to 
the rotation axis the geometric explanation of the non-detection
becomes less plausible.
This issue can be clarified by observing the
same target at multiple epochs.

\subsection{Flaring emission}

If the emission is sporadic, it may have been missed
because of our time sampling of 16~s. 
However, this is not the most likely explanation 
because even in the case 
of Jupiter's flares, the duty cycle within 
the time-frequency space is between 15 and 30\%. The corresponding
dilution factor is only between 3 and 6, and is not crucial given the large 
uncertainties in the other parameters of the system. 

\subsection{Emission flux}

For the three mechanisms considered: i.e., kinetic 
emission, magnetic emission, and coronal mass ejection, the corresponding 
radio flux density estimates should reach 0.4, 900, and 30~mJy, 
respectively (Greissmeier 2007). 
The magnetic emission is the favored alternative but the emission is expected 
to peak at just a few MHz, i.e., at frequencies much lower than those covered
in the present observations.

Nonetheless, there is another possible scenario.
For an intense magnetic field (of up to 40\,G; Moutou et al.\ 2007), 
an active chromosphere (Boisse et al.\ 2009), and a rapid rotation
(11.7 days period; H\'ebrard \& Lecavelier des Etangs 2006; Croll et 
al.\ 2007), the stellar magnetosphere can extend beyond the orbit of the 
extrasolar planet whose semi-major axis is only 0.03\,AU (see discussion in 
Jardine \& Cameron 2008). 
Assuming that the emission is produced by an
interaction between the planetary magnetosphere and the stellar corona 
in which it is embedded, and using Eq.~17 and the numerical values from the 
model of Jardine \& Cameron (2008), 
with a 10\% efficiency of conversion of the power of accelerated 
electrons into radio emission and an emission beam solid angle of 1.6\,sr,
the predicted radio flux from HD189733\,b is about 15\,mJy.  
With this model and assuming that our non-detection of emission 
from HD189733\,b is caused by its low intrinsic intensity, 
the upper limits of 2\,mJy at 
244\,MHz and 0.16\,mJy at 614\,MHz can be translated into 
upper limits for the stellar coronal 
density of 0.36 and 0.1~times the solar coronal density, respectively. 

\subsection{Emission frequency}

The principal mechanism advocated for radio emission is the electron-cyclotron maser radiation. 
It occurs at the local gyrofrequency $f_g$ given by, 
$f_g=2.8\left({B_p}/{\rm 1\,G}\right) {\rm MHz}$,
where $B_p$ is the planet's magnetic field strength. 
The non-detection could then be attributed to a weak planetary
magnetic field, such that the gyrofrequency falls below our observation 
frequencies. The observed frequencies of 244 and 614~MHz would then correspond to 
planetary magnetic fields of 85~G and 220~G, respectively. We recall that the 
Jovian magnetic field strength estimated from the radio frequency cut-off of 
its cyclotron emission is around 40~G. 

Finally, cyclotron maser emission can also be quenched by a 
too high plasma frequency ($f_{pe}$) in the source region (Zarka 2007). 
With a 40~G magnetic field, the condition for quenching ($f_{pe}$ larger 
than a tenth of the cyclotron frequency) implies an electron density larger 
than about 1.5$\times$10$^{6}$~cm$^{-3}$ in the low stellar corona, 
a condition which is not implausible.

\section{Conclusion}

In summary, our radio observations of HD189733\,b have provided very tight upper limits
at 244~MHz and 614~MHz, which are below some predictions reported in the literature. 
The non-detection could be attributed to the inadequate time sampling rate of the observation, beam focusing, or 
intrinsic emission power being lower than theoretical predictions.
The frequencies of our observations would require strong planetary magnetic field,
and so we favor the scenario of a low magnetic field as an explanation of 
our non-detection.
Better prospects are clearly offered by observations at lower frequencies, 
which will become feasible with UTR2, LOFAR (planned in short-term future), 
and SKA (long-term future).

\begin{acknowledgements}
We thank the anonymous referee for the helpful comments.
We thank the staff of the GMRT who have made these observations possible. 
Special thanks are due to J. P. Kodilkar for his help in executing the
observations.
GMRT is run by the National Centre for Radio Astrophysics (NCRA) 
of the Tata Institute of Fundamental Research (TIFR).
P.Z. activities in radio search for exoplanets are partly supported by
ANR program NT05-1\_42530 "Radio-Exopla".
\end{acknowledgements}


\begin{thebibliography}{}

\bibitem[2006]{Bakos et al.2006}Bakos, G.~{\'A}., Knutson, H., Pont, F., et 
al.\ 2006, \apj, 650, 1160 

\bibitem[2000]{Bastian et al.2000}Bastian, T.~S., Dulk, G.~A., 
\& Leblanc, Y.\ 2000, \apj, 545, 1058 

\bibitem[2009]{Boisse et al.2009}Boisse, I., Moutou, C., Vidal-Madjar, A., 
et al.\ 2009, \aap, 495, 959 

\bibitem[2005]{Bouchy et al.2005}Bouchy, F., Udry, S., Mayor, M., et al.\ 
2005, \aap, 444, L15 

\bibitem[2005]{Charbonneau et al.2005}Charbonneau, D., Allen, L.~E., 
Megeath, S.~T., et al.\ 2005, \apj, 626, 523 

\bibitem[2008]{Charb2008} 
 Charbonneau, D., Knutson, H.~A., Barman, T., et al.\ 2008, ApJ, 686, 1341 

\bibitem[2007]{Croll et al.2007}Croll, B., Matthews, J.~M., Rowe, J.~F., et 
al.\ 2007, \apj, 671, 2129 

\bibitem[2005]{Deming et al.2005}Deming, D., Seager, S., Richardson, L.~J., 
\& Harrington, J.\ 2005, \nat, 434, 740 

\bibitem[2006]{Deming et al.2006}Deming, D., Harrington, J., Seager, S., 
\& Richardson, L.~J.\ 2006, \apj, 644, 560 

\bibitem[2009]{Desert et al.2009}Desert, J.-M., Lecavelier des Etangs, A., 
Hebrard, G., et al.\ 2009, arXiv:0903.3405 

\bibitem[2007]{George et al.2007}George, S.~J., 
\& Stevens, I.~R.\ 2007, \mnras, 382, 455 

\bibitem[2007]{Griessmeier2007}Griessmeier, J.-M., Zarka, P., Spreeuw, H.\ 2007
\aap, 475, 359

\bibitem[2009]{Grillmair}Grillmair, C.~J., Burrows, A., Charbonneau, D., et al.\ 2009, 
Nature, 456, 767  

\bibitem[2006]{Hebrard et al.2006}H{\'e}brard, G., 
\& Lecavelier Des Etangs, A.\ 2006, \aap, 445, 341 

\bibitem[2008]{Jardine et al.2008}Jardine, M., 
\& Cameron, A.~C.\ 2008, \aap, 490, 843 

\bibitem[2009]{Knutson et al.2009}Knutson, H.~A., Charbonneau, D., Cowan, 
N.~B., et al.\ 2009, \apj, 690, 822 
\bibitem[2008]{Lecavelier Des Etangs et al.2008}Lecavelier Des Etangs, A., 
Pont, F., Vidal-Madjar, A., \& Sing, D.\ 2008, \aap, 481, L83 

\bibitem[2007]{Lazio et al.2007}Lazio, T.~J.~W., 
\& Farrell, W.~M.\ 2007, \apj, 668, 1182 

\bibitem[2009]{Lazio et al.2009}Lazio, J., Bastian, T., Bryden, G., 
et al.\ 2009, arXiv:0903.0873 

\bibitem[2008]{Lecav2008}Lecavelier des Etangs, A.,
Pont, F., Vidal-Madjar, A., \& Sing, D.\ 2008, \aap , 481, L83

\bibitem[2007]{Moutou et al.2007}Moutou, C., Donati, J.-F., Savalle, R., et 
al.\ 2007, \aap, 473, 651 

\bibitem[2007]{Pont et al.2007}Pont, F., Gilliland, R.~L., Moutou, C., et 
al.\ 2007, \aap, 476, 1347 

\bibitem[2008]{Pont et al.2008}Pont, F., Knutson, H., Gilliland, R.~L., 
Moutou, C., \& Charbonneau, D.\ 2008, \mnras, 385, 109 

\bibitem[2008]{Redfield et al.2008}Redfield, S., Endl, M., Cochran, W.~D., 
\& Koesterke, L.\ 2008, \apjl, 673, L87 

\bibitem[2004]{Ryabov et al.2004}Ryabov, V.~B., Zarka, P., 
\& Ryabov, B.~P.\ 2004, \planss, 52, 1479 

\bibitem[2009]{Sirothia2009}Sirothia S.~K.\ 2009, MNRAS submitted

\bibitem[2009]{Smith et al.2009}Smith, A.~M.~S., Cameron, A.~C., Greaves, 
J., et al.\ 2009, \mnras, 395, 335 

\bibitem[1990]{Swarup90}Swarup, G.\ 1990, Indian Journal of Radio and Space Physics, 19, 493

\bibitem[2007]{Winn et al.2007}Winn, J.~N., Holman, M.~J., Henry, G.~W., et 
al.\ 2007, \aj, 133, 1828 

\bibitem[2005]{Winterhalter et al.2005}Winterhalter, D., Majid, W., Kuiper, 
T., et al.\ 2005, Bulletin of the American Astronomical Society, 37, 1292 

\bibitem[2007]{Zarka2007}Zarka, P.\ 2007, \planss, 55, 598 

\bibitem[2001]{Zarka et al.2001}Zarka, P., Treumann, R.~A., Ryabov, B.~P., 
\& Ryabov, V.~B.\ 2001, \apss, 277, 293 

\end{thebibliography}
\end{document}